\documentclass[aps, prd,linenumbers, twocolumn,superscriptaddress,preprintnumbers]{mnras}

\usepackage{natbib}
\include{notations}
\usepackage{natbib}

\include{notations}

\usepackage{amsmath,amssymb,latexsym,times}

\usepackage{slashed}
\usepackage{graphicx}
\usepackage{grffile}
\usepackage[normalem]{ulem}
\usepackage{dcolumn}
\usepackage[dvipsnames]{xcolor}
\usepackage{booktabs}
\usepackage{amssymb}
\usepackage{slashed}
\usepackage{todonotes}
\usepackage{hyperref}
 \hypersetup{
     colorlinks=true,
     linkcolor=aiiro,
     citecolor = aiiro,      
     urlcolor=aiiro,
     }
\usepackage{comment}
\usepackage{footnote}
\usepackage{newtxtext,newtxmath}

\usepackage[T1]{fontenc}
\usepackage{graphicx}	
\usepackage{amsmath}	
\usepackage{amssymb}	
\usepackage{slashed}
\usepackage[normalem]{ulem}
\usepackage{xspace}
\definecolor{aiiro}{HTML}{105779}
\usepackage{lineno}
\usepackage{parskip}

\def\Cref#1{Chapter~\ref{#1}\xspace}

\newcommand{\nbody}{$n$-body}

\definecolor{ircol}{HTML}{B72654}
\usepackage{lipsum}

\usepackage{anyfontsize}

\graphicspath{{./}{figures/}}

\begin{document}

\title{Detecting deviations from Gaussianity in high-redshift CMB lensing maps}

\date{\today}

\author[Z. Zhang, Y. Omori, and C. Chang]{Zhuoqi (Jackie) Zhang$^{\star,1}$, Yuuki Omori$^{1,2}$, and Chihway Chang$^{1,2}$
\\ \\
$^{1}$ Department of Astronomy and Astrophysics, University of Chicago, Chicago, IL 60637, USA\\
$^{2}$ Kavli Institute for Cosmological Physics, University of Chicago, Chicago, IL 60637, USA\\
$^{\star}$ E-mail: zhuoqizhang@uchicago.edu}

\maketitle

\begin{abstract}
While the probability density function (PDF) of the cosmic microwave background (CMB) convergence field approximately follows a Gaussian distribution, primordial non-Gaussianities and small contributions from structures at low redshifts make the overall distribution slightly non-Gaussian. Some of the late-time component can be modelled using the distribution of galaxies and subtracted off from the original CMB lensing map to produce a map of matter distribution at high redshifts. Using this high-redshift mass map, we are able to directly study the early phases of structure formation. In this work, we forecast the detectability of signatures of non-Gaussianity due to nonlinear structure formation at $z>1.2$. Assuming the optimal case of no systematics, we find that it is challenging to detect such signatures in current surveys, but future experiments such as the deep field of CMB-S4 will be able to make detections of $\sim7\sigma$. \\ \hfill\\ 
\textbf{Key Words:} gravitational lensing: weak, cosmology: cosmic background radiation \\
\end{abstract}
\vspace{0.2in}
\section{Introduction}

Photons from the cosmic microwave background (CMB) are gravitationally lensed as they pass through the large-scale structure (LSS) between the last scattering surface and us the observer. The overall amplitude and the characteristics of this lensing effect are sensitive to the standard cosmological parameters such as $\sigma_{8}$ and $\Omega_{\rm m}$, as well as the sum of neutrino masses $\Sigma {m}_{\nu}$, and curvature of the Universe $\Omega_{\rm K}$ \citep{planck2018}. To date, these information have mainly been extracted through the auto-spectrum of the reconstructed CMB lensing map \citep{planck2013,planck2015,planck2018,simard2018,bianchini2020}. as well as through measurements of cross-correlations with the LSS \citep{DES-Y1-5x2,krolewski2021,robertson2021,white2022,DES-Y3-5x2}.

The probability density function (PDF) of the CMB lensing field is known to be nearly Gaussian. This is due to two factors: first, since the source plane of the CMB lensing field is at $z\sim 1100$, the lensing effect is integrated over many independent LSS volumes, which tends to make the overall field more Gaussian. Second, the lensing signal is dominated by information sourced by LSS at $z\sim2$, where the effects from the nonlinear structure formation are much more subtle than at lower redshifts \citep{hanson2010}. On the other hand, several factors do contribute to non-Gaussian signatures. Although the amplitude is small, the low-redshift LSS (which is significantly non-Gaussian) contributes to the overall signal and leaves the CMB lensing field slightly non-Gaussian. Furthermore, primordial non-Gaussianity generated from certain inflationary models can introduce non-Gaussian structures at high-redshift before nonlinear structure starts to form. Previous works have shown that summary statistics sensitive to these non-Gaussianities could be exploited to extract information on late-time structure formation beyond what the power spectra could provide \citep[e.g.][]{pdf-peak-count,planck2013pdf}. 

While it is possible to extract cosmological information from the non-Gaussianities in CMB lensing maps, one may argue that other observational probes that are more sensitive to low-redshift structures, such as galaxy weak lensing, are more efficient at capturing these signals. Nonetheless, there are several advantages of using the CMB lensing map for non-Gaussianity studies. First, galaxy weak lensing maps are subject to a number of systematics that are unique to galaxies, such as intrinsic alignment, photometric redshift and calibration of shear measurements. Having an additional complementary probe for non-Gaussianity such as CMB lensing is therefore highly beneficial. Furthermore, the lensing kernel for CMB peaks at a different redshift than the weak lensing kernel, allowing us to push this study to a different (higher) redshift range. Finally, measuring the signal at high redshift where non-Gaussianities from structure formation has not yet kicked in could potentially be beneficial for studying primordial non-Gaussianity. 

The main challenge in detecting signatures of non-Gaussianity at high redshifts is that the low-redshift structures imprint signals that dominate over those from high-redshift structures. That is, if we want to isolate the contributions from high redshifts, some additional procedure is needed to remove the low-redshift contribution.

In recent years, interests in \textit{delensing} have grown within the CMB community: a procedure of reversing the distortions in the temperature and polarisation maps due to lensing, using various tracers of the gravitational potential (such as galaxies and the cosmic infrared background). Delensing is primarily used to remove the lensing $B$-modes contributions from the measurements of $B$-modes \citep{bkspt2021} or used to sharpen the acoustic peaks in the measured $T$/$E$ power spectra \citep{planck2018}. However, the same technique can be employed to remove the low-redshift portion of the CMB lensing map. In \cite{qu2022}, the authors point out that this high-redshift CMB lensing map is useful in extracting cosmological information, since it is less affected by astrophysical systematic effects (such as baryonic feedback effects), as well as nonlinear structure formation.

In this paper, we forecast the detectability of non-Gaussianity due to nonlinear structure formation in CMB lensing maps at $z>1.2$. We approach this by first generating simulated CMB lensing maps for various CMB experiments, removing the low-redshift contribution using a mock galaxy survey, and measuring the PDF of the high-redshift lensing map. Our methodology is similar to \cite{qu2022}, but the focus of our work is the non-Gaussian signal in the high-redshift Universe.

In this work, we simulate the following ongoing/near-future CMB experiments: the third-generation South Pole Telescope survey (SPT-3G; \citealt{spt-3g}), Simon's Observatory (SO; \citealt{simons}), and both the wide and deep surveys of the CMB-S4 experiment \citep{cmb-s4}. For the low-redshift tracer, we use the galaxy samples from the Vera C. Rubin Observatory Legacy Survey of Space Time (LSST) Year 1 dataset. 

This paper is structured as follows: We first outline the theoretical modelling and our procedure to remove the low-redshift contribution from the CMB lensing maps in Section~\ref{sec:background}. We then describe the simulations we use to perform our tests in Section~\ref{sec:input}. We present the main results in Section~\ref{sec:results}, and finally conclude in Section~\ref{sec:conclusion}. Throughout the paper, we adopt the cosmological parameters from \cite{planck2013cosmo} with $h=0.678$, $\Omega_{\rm m}=0.307$, $\Omega_{\Lambda}=0.693$, $\Omega_{\rm b}=0.048$, $n_{2}=0.96$, and $\sigma_{8}=0.818$.

\section{Modeling and Analysis}\label{sec:background}

\subsection{Auto- and cross- spectrum}\label{sec:cross_correlation}

We first describe how the auto- and cross-spectrum between galaxies and CMB lensing are computed, as they are required in constructing the low-redshift lensing template. The auto- and cross-power spectra of two fields ($\alpha$ and $\beta$) are given by:
\begin{align}
    C_\ell^{\alpha,\beta}=\frac{2}{\pi}\int_0^\infty dk\int_0^{\chi_\text{max}} d\chi \int_0^{\chi'_\text{max}} d\chi'k^2q^\alpha(\chi)j_\ell(k\chi) \notag\\ q^\beta(\chi')j_\ell(k\chi')P(k),
\end{align}
where $\alpha,\beta$ can be galaxy density $\delta$ or CMB lensing $\kappa$, $\chi$ is the comoving radial distance, $P$ is the matter power spectrum\footnote{We choose to use the nonlinear halofit model from \citealt{halofit-takahashi}, except in the calculation of the filter function in Equation~\eqref{eq:filter}}, and $j_\ell$ is the spherical Bessel function; the upper bound of integrals, $\chi_{\rm max}$ and $\chi'_{\rm max}$ are either the maximum comoving distance of the galaxy sample (for the case of $\delta\delta$ or $\delta\kappa$ correlation) or the comoving distance to the surface of last scattering (for the case of $\kappa\kappa$). $q^\alpha$, $q^\beta$ are the tracer kernels. For CMB lensing, the kernel is: 
\begin{equation}
    q^\kappa(\chi)=\frac{3\Omega_{\rm m}H_0^2}{2c^2}\frac{\chi}{a(\chi)}\frac{\chi^*-\chi}{\chi^*},
\end{equation}
where $\Omega_{\rm m}$ is the matter density, $H_0$ is the Hubble constant today, and $a$ is the cosmological scale factor. For a specific galaxy tomographic redshift bin, the kernel is given by:
\begin{equation}
    q^{\delta}(\chi)=b^{\delta}n^{\delta}(z(\chi))\frac{dz}{d\chi},
\end{equation}
where $n^{\delta}$ stands for the galaxy density as a function of redshift and is normalized to one, and $b^{\delta}$ stands for the linear galaxy bias of the tomographic bin, which is assumed to be a constant.

\subsection{Removal of low-$z$ lensing}\label{sec:delens}

To calculate the low-redshift contribution of the total CMB lensing signal, we first estimate the underlying matter field by observing the galaxy density field and ``undoing" the galaxy kernel. The CMB lensing kernel is then applied to the estimated  matter density field to produce a template of the low-redshift CMB lensing map.

Formally, this procedure can be written as \citep{planck2018}:
\begin{equation}
    \kappa_{\ell m}^{{\rm low-}z}=\sum_i w^i\frac{\sqrt{C_\ell^{\kappa\kappa}}}{\sqrt{C_\ell^{\delta_i\delta_i}}}\Delta_{\ell m}^i\label{eq:delens_alm},
\end{equation}
where $\Delta^i_{\ell m}$ are the $a_{\ell }m$ of the $i$-th redshift bin galaxy density map, and $w^i$ is the weight for the $i$-th redshift bin, which is given by
\begin{equation}
    w^i=\sum_j \rho^{\kappa}_{j}[(\boldsymbol\rho^{\delta})^{-1}]_{ij},
\end{equation}
where $\rho$ is the correlation between different redshift bins of the galaxy density field and CMB lensing:\footnote{To compute $\rho_{ij}^{\delta}$ and $\rho_i^\kappa$, shot-noise should be included in the galaxy-galaxy power spectra $C_\ell^{\delta_i\delta_j}$. However, one does not need $C_\ell^{\kappa\kappa}$ to obtain the high-redshift CMB lensing map since the $C_\ell^{\kappa\kappa}$ term in $\rho_i^\kappa$ cancels with that in Equation~\eqref{eq:delens_alm} and leaving Equation~\eqref{eq:delens_alm} independent of $C_\ell^{\kappa\kappa}$.}  
\begin{align}
    \rho^{\delta}_{ij}&=C_\ell^{\delta_i\delta_j}/
\sqrt{C_\ell^{\delta_i\delta_i} C_\ell^{\delta_j\delta_j}},\\
\rho^{\kappa}_{i}&=C_\ell^{\delta_i\kappa}/\sqrt{C_\ell^{\delta_i\delta_i}C_\ell^{\kappa\kappa}}.\label{eq:rho_ik}
\end{align}
Once we create the low-redshift template, we subtract it from the full CMB lensing map to isolate the high-redshift CMB lensing map:
\begin{equation}
\kappa^{{\rm high}-z}_{\ell m }=\kappa^{\rm full}_{\ell m }-\kappa^{{\rm low}-z}_{\ell m }.\label{eq:delensed_kappa}
\end{equation}
We also note that an analytical expression of the high-redshift-only CMB lensing spectrum $C_{\ell}^{\kappa\kappa,{\rm high-}z}$ can be derived using Equations ~\eqref{eq:delens_alm}--\eqref{eq:rho_ik} \citep{Yu-delens}:
\begin{equation}
    C_{\ell}^{\kappa\kappa,{\rm high-}z}= C_{\ell}^{\kappa\kappa}-\sum_{i,j}\rho^{\kappa}_{i}[(\boldsymbol\rho^{\delta})^{-1}]_{ij}\rho^{\kappa}_{j}C_{\ell}^{\kappa\kappa}.\label{eq:delens_cl}
\end{equation}
We verify that our methodology recovers the expected high-redshift CMB lensing map using a simulation suite that contains both CMB lensing and LSS observables (see Appendix~\ref{sec:validation}).
\begin{figure}
    \centering
    \includegraphics[width=\linewidth]{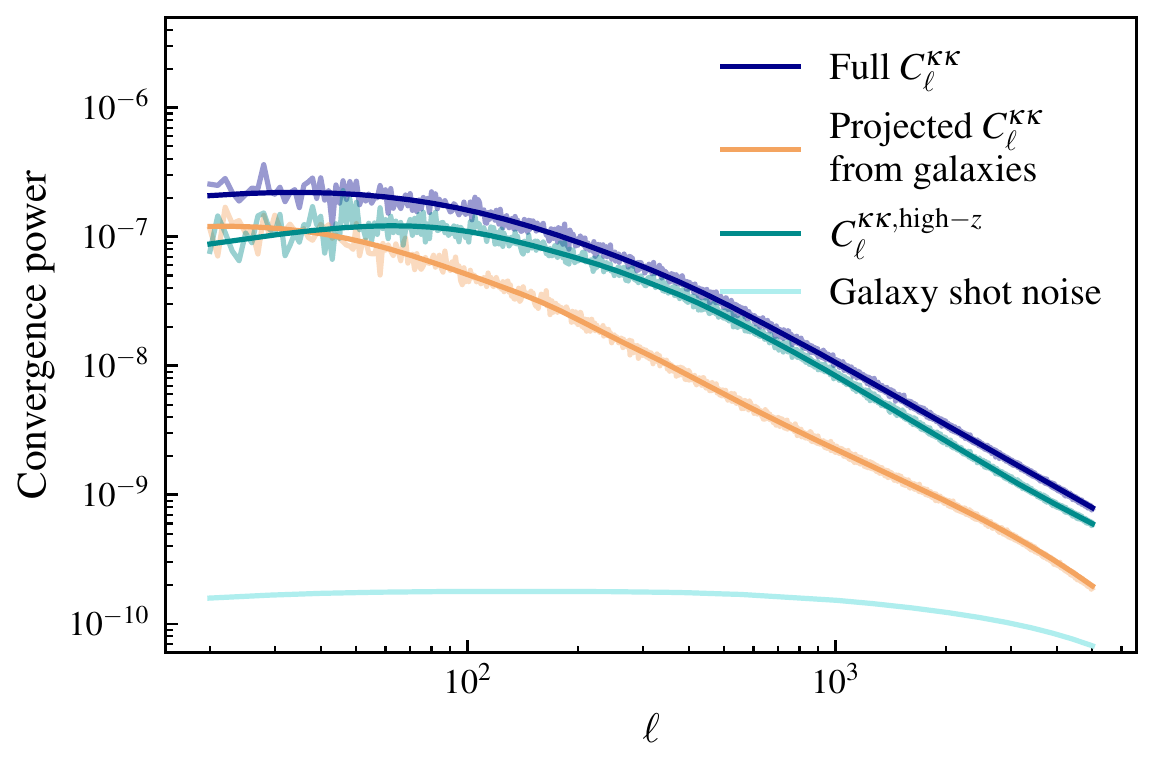}
    \caption{
    Plot showing the full CMB lensing spectrum (dark blue), the low-redshift component of the lensing spectrum estimated from LSST-Y1 galaxies (orange), and their difference (i.e. high-redshift CMB lensing spectrum; teal).
    The solid lines are computed using Equation~\eqref{eq:delens_cl}, and the transparent lines are measured power spectra from a simulation. $C_\ell^{\kappa\kappa}$,  $C_\ell^{\kappa\kappa,{\rm high-}z}$, and projected $C_\ell^{\kappa\kappa}$ are defined according to Equation~\eqref{eq:delensed_kappa}; the galaxy shot noise curve shows the contribution of galaxy shot noise to $C_\ell^{\kappa\kappa,{\rm high-}z}$, computed by substituting $\Delta_{\ell m}^i$ in Equation~\eqref{eq:delens_alm} with the galaxy shot noise map.} 
    \label{fig:delens_ideal}
\end{figure}
In Figure \ref{fig:delens_ideal}, we show the full CMB lensing spectrum, low-redshift CMB lensing derived from galaxies, the high-redshift CMB lensing signal, and the effective\footnote{The effective shot-noise is computed by passing a shot-noise map generated using the expected number density of galaxies through Equations 4-7. } shot noise of the galaxy sample used.

\subsection{Filtering}\label{sec:filter}
Since we expect most of the non-Gaussian signal to be associated with nonlinear structure evolution, we apply the following filter to the constructed high-redshift CMB lensing maps to focus on the scales where we expect the highest signal:
\begin{equation}
    f_{\ell}=\frac{C_{\ell,{\rm nonlin}}^{\kappa\kappa,{\rm high-}z}-C_{\ell,{\rm lin}}^{\kappa\kappa,{\rm high-}z}}{C_{\ell,{\rm nonlin}}^{\kappa\kappa,{\rm high-}z}-C_{\ell,{\rm lin}}^{\kappa\kappa,{\rm high-}z}+N_\ell^{\kappa\kappa}},\label{eq:filter}
\end{equation}
where the subscript ``lin"/``nonlin" denotes whether the spectrum is computed using the linear or nonlinear matter power spectrum, and $N_{\ell}^{\kappa\kappa}$ is the CMB lensing noise spectrum.\footnote{While this filter is cosmology-dependent, we expect that a slight offset in the cosmology assumed would only cause minor modifications to our forecast results.} We apply this filter to $\kappa_{\ell m}^{{\rm high}-z}$ computed in Equation \eqref{eq:delensed_kappa}. 

\begin{figure}
    \centering
    \includegraphics[width=\linewidth]{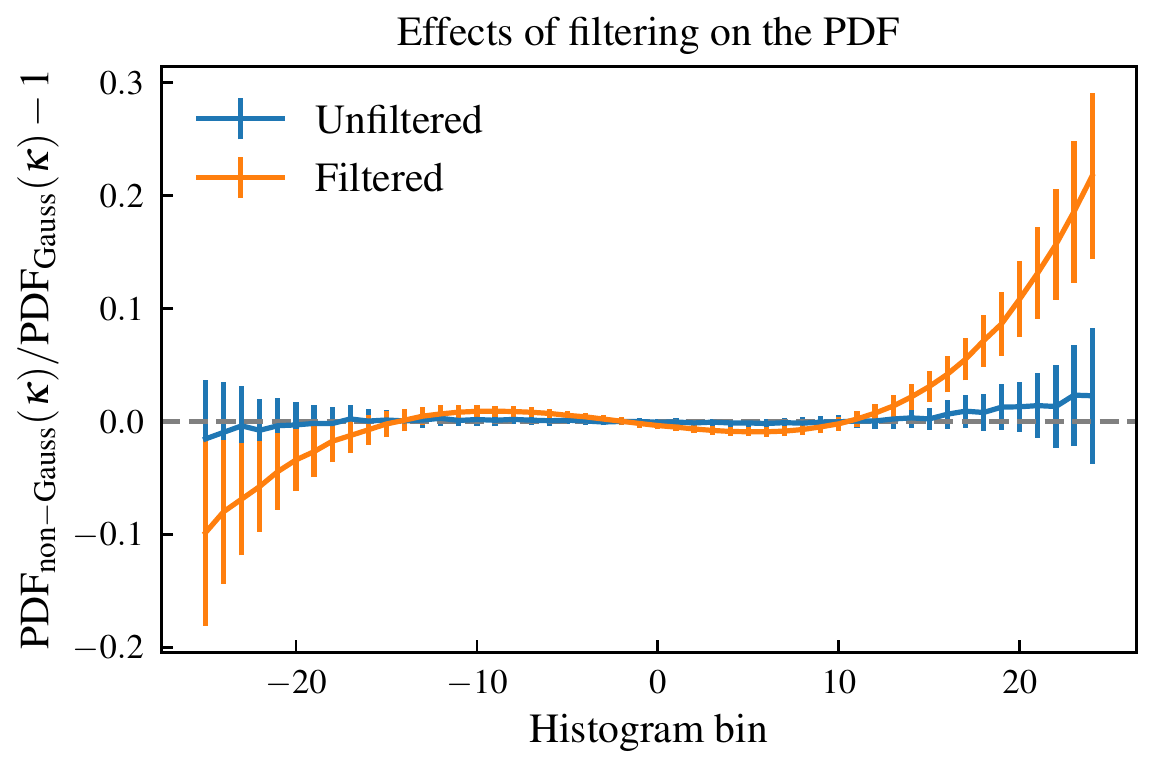}
    \caption{The fractional difference between the PDFs of non-Gaussian and Gaussian simulations (see Section~\ref{sec:input}) with and without the filter defined in Equation \eqref{eq:filter} applied.}
    \label{fig:filter}
\end{figure}

We compare the fractional difference between the PDFs from simulated non-Gaussian and Gaussian maps in Figure~\ref{fig:filter}. Without applying this filter, the PDF of the non-Gaussian map is indistinguishable from that of a Gaussian field. However, after applying the filter, we find a clear and statistically significant difference between the PDFs of the two fields.

\subsection{Detection of deviations from Gaussianity}\label{sec:measure}

In this work, we focus on measuring the PDFs of the pixel values of the high-redshift CMB lensing convergence map to detect deviations from Gaussianity. While the PDF is not the most sensitive summary statistic to capture non-Gaussianity of a specific type, the foremost advantage is its simplicity: the PDF is easy and fast to measure since it is, in practice, simply the histograms of pixels values of the CMB lensing map.

In linear theory, the histogram of the pixel values are statistically consistent with a Gaussian distribution. In the late Universe, the overdensities tend to grow at a different rate compared to underdensities. As a result, the PDF of the density field (and hence the lensing field) becomes skewed towards positive values, resulting in a non-Gaussian distribution. 

We quantify the amount of non-Gaussianity in a given map by measuring the $\chi^2$ between its PDF and that of a Gaussian random field:
\begin{equation}
    \chi^2=({\bf d_{\rm NG}-\bar{d}}_\text{G})^{\rm T}{\rm Cov}^{-1}({\bf d_{\rm NG}-\bar{d}}_\text{G}),
\end{equation}
where $\bf d_{\rm NG}$ is the observed data vector, i.e. PDF measured from data, which we will simulate using an $N$-body simulation (see Section~\ref{sec:datavecs}); ${\bf\bar d}_\text{G}$ is the mean PDF predicted using Gaussian simulations, and $\text{Cov}$ is the covariance matrix for the data vector, also obtained from those simulations described in Section~\ref{sec:gaussian_sim}. Finally, we adopt the definition of signal-to-noise from \citet{secco-s2n} and use:
\begin{equation}
    S/N=\sqrt{\chi^2-\nu},\label{eq:n-sigma}
\end{equation}
where $\nu$ is the number of degrees of freedom, i.e. the length of the data vector. To reduce statistical uncertainty, we will generate multiple observed data vectors and report the mean and spread of $\chi^2$s and $S/N$s measured from individual realisations.

\section{Simulations}\label{sec:input}

In this section, we describe the assumed noise levels and sky coverage for the various ongoing and upcoming CMB experiments (SPT-3G, SO, CMB-S4). We also describe the number densities and redshift distributions assumed for the LSST Year-1 galaxy sample. From these setups, we generate two sets of simulated CMB lensing maps: the ``Gaussian'' maps that serve as a reference and the ``non-Gaussian'' maps that contain realistic non-Gaussian information from the LSS. The two sets of simulations are described in Sections~\ref{sec:gaussian_sim} and \ref{sec:datavecs}.

\subsection{CMB experiments}\label{sec:CMB_surveys}

\begin{figure}
    \centering
    \includegraphics[width=\linewidth]{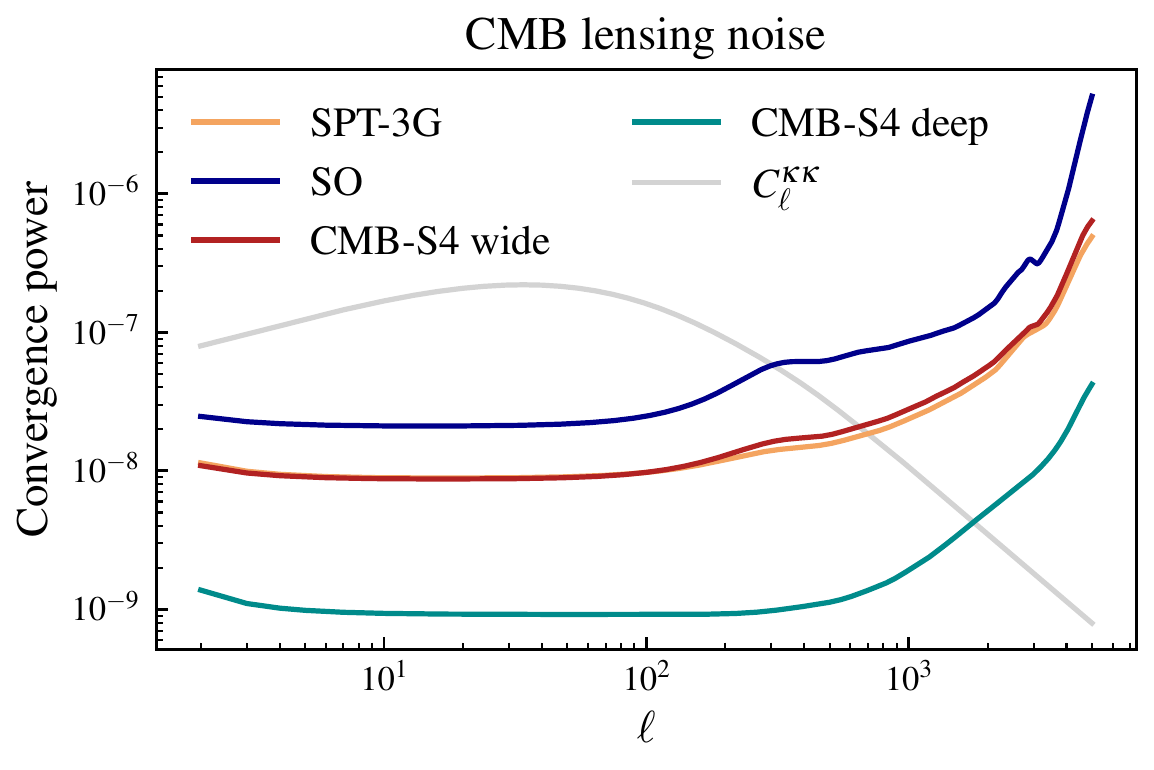}
    \caption{CMB lensing noise power spectra for the surveys mentioned in Section~\ref{sec:CMB_surveys}.}
    \label{fig:CMB_noise}
\end{figure}

We simulate data from several on-going and future CMB experiments. The basic properties and characteristics of the data from these experiments are listed below. Figure~\ref{fig:CMB_noise} shows the lensing noise power spectra for the four surveys described below. 

\begin{enumerate}
\item \textbf{SPT-3G} is a 10-meter telescope operating at the geographical South Pole. It is currently making deep observations of the CMB over 1500 deg$^2$ or 3.5\% of the sky in the Southern Hemisphere.
\item \textbf{SO} is a 6-meter telescope located at the Atacama Desert of Northern Chile, scheduled to start observing in 2023. Although the noise level in SO is projected to be higher than SPT-3G, SO will observe a much larger area, roughly 40\% of the total sky. For this work, we will use the projected \textit{goal} lensing noise power spectrum. 
\item \textbf{CMB-S4} is the next-generation ground based CMB experiment that is set to begin observing in 2027 . It is planned to combine several telescopes in the Atacama and the South Pole to make exquisite maps of the millimeter sky. We adopt both their ``Wide" configuration (covering 40\% of the sky) and their ``deep configuration" (covering 3.5\% of the sky).
\end{enumerate}

\subsection{Galaxy samples}
\label{sec:galaxy}
\begin{figure}
    \centering
    \includegraphics[width=\linewidth]{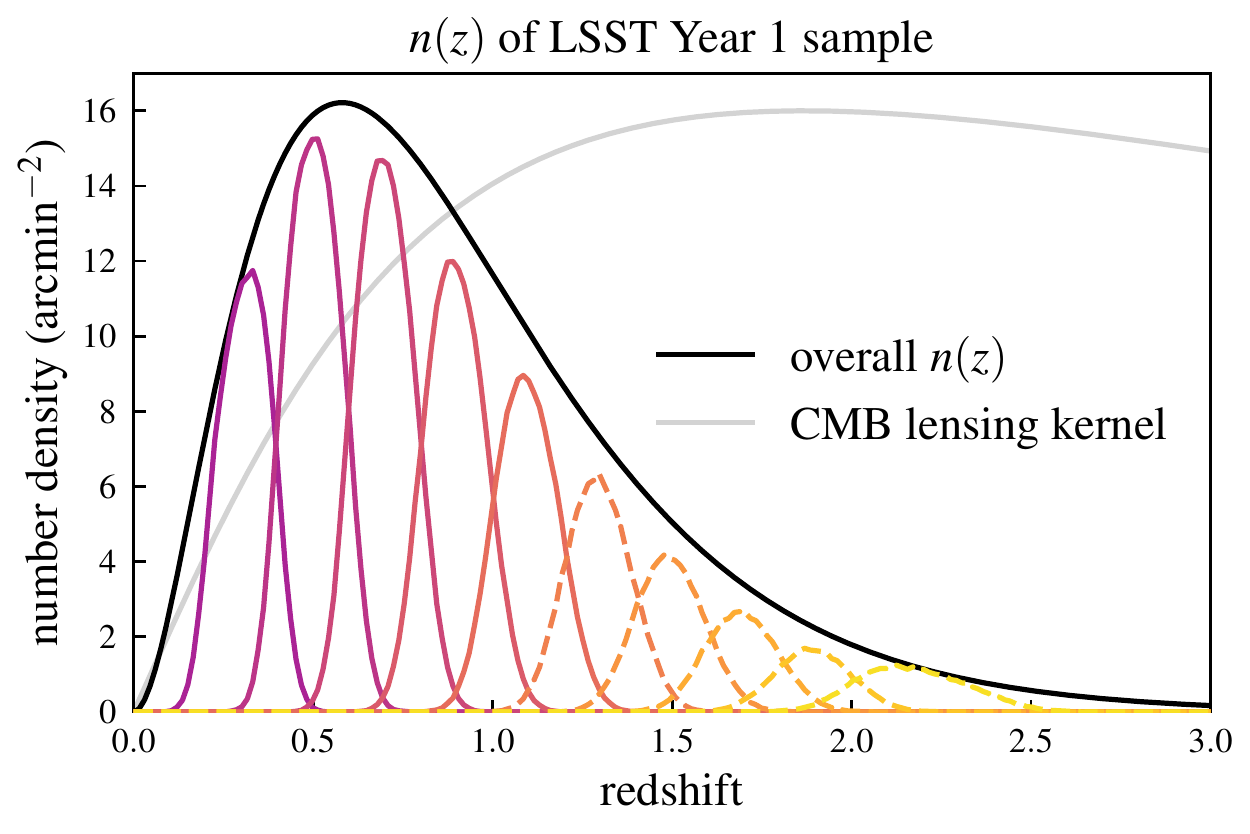}
    \caption{The $n(z)$ for LSST Year 1 sample. The black envelope shows the overall $n(z)$. The solid lines show the fiducial samples LSST DESC SRD assumes. The dashed lines show the additional high-redshift bins we assume for testing redshift dependence. The grey line shows the CMB lensing kernel (up to a normalization constant).}
    \label{fig:nz}
\end{figure}

We will use the specifications of galaxy samples defined in the LSST Dark Energy Science Collaboration (DESC) Science Requirement Document \citep[DESC SRD,][]{SRD2018}\footnote{Estimates in the SRD are \textit{requirements} rather than (optimistic) projections. These estimates are used for LSST official forecasts.}. In particular, we will use the first year (Y1) lens galaxy sample specified therein for this analysis. LSST is a ground based galaxy survey that is planned to cover roughly 12,000 deg$^2$ of the sky that approximately subsumes footprints of all the CMB experiments described above. The LSST survey will start observing $\sim$2024 and the Y1 data is expected to be released publicly in $\sim2025$. According to the DESC SRD, the overall redshift distribution of the galaxy sample $n(z)$ is assumed to take the form 
\begin{equation}
    dN/dz\propto z^2\exp\left[-(z/z_0)^\alpha\right],
\end{equation}
with $(z_0, \alpha)=(0.26,0.94)$, and with a total integrated number density of 18 arcmin$^{-2}$. The photo-$z$ uncertainty is specified to be $\sigma_z=0.03(1+z)$. The full galaxy sample is divided into 5 redshift bins between $0.2 <z< 1.2$, each with $\Delta z=0.2$. In addition, we create an additional 5 redshift bins ranging from $1.2<z<2.2$, with the same bin widths to investigate optimistic scenarios. The redshift
distribution of our galaxy sample is shown in Figure~\ref{fig:nz}. The galaxy bias is assumed to be linear and is approximated as a constant throughout each redshift bin. The dependence of galaxy bias on redshift bin is assumed to take the following form: 
\begin{equation}
    b(z) = 1.05/G(z),
\end{equation}
where $G(z)$ is the normalized growth factor (with $G(0)=1$), and for each redshift bin, the bias is evaluated at the mean redshift. 

We note that the LSST-Y1 sample, as well as many other surveys, do not use galaxies at $z<0.2$. Therefore, with our methodology, we cannot remove the lensing contributions from $z<0.2$. However, since the CMB lensing is very insensitive to those redshifts, the non-Gaussian signals from this redshift range is small compared to those from the higher-redshift ranges.

\subsection{Gaussian simulations}\label{sec:gaussian_sim}

We generate Gaussian realisations of the CMB lensing and galaxy density maps for two purposes: (a) to obtain the reference PDF data vector against which we compare our non-Gaussian PDF, and (b) to estimate the covariance of the measured PDF data vectors. For the latter, one would ideally use multiple realisations of the non-Gaussian simulations (see Section~\ref{sec:datavecs}). However, generating $\mathcal{O}(\sim1000)$ fully ray-traced non-Gaussian simulations at the resolution and sky area we require is computationally unfeasible. Therefore, we use the covariance estimated from cheaper Gaussian realisations instead. Given that the CMB lensing field is nearly Gaussian, we make the assumption that the difference between the covariance matrices generated from Gaussian vs. non-Gaussian simulations is sufficiently small for the purpose of this work.  

We first compute the auto and cross spectra among galaxy redshift bins and CMB lensing using \texttt{CAMB} as described in Section~\ref{sec:cross_correlation} and then generate noiseless correlated Gaussian realizations of galaxy density and CMB lensing maps accordingly.\footnote{Correlated maps and spherical harmonics can be generated using the method described in Appendix A of \cite{correlated_map}. For our case, it is more convenient to generate the spherical harmonics, and the procedure is described by Equations (A4) and (A8) in \cite{correlated_map}.} Next, we add shot noise and CMB noise respectively, where the shot noise power spectrum is given by $1/N^{\delta_i}$, where $N^{\delta_i}$ is the galaxy number density of the $i$-th bin in units galaxies per steradian as specified in Section~\ref{sec:galaxy}, and the CMB lensing noise power spectra is shown in Figure~\ref{fig:CMB_noise}. We remove the low- redshift contribution to the CMB lensing map as described in Section~\ref{sec:background} using these galaxy density maps and produce a  histogram of the high-redshift lensing map to obtain a binned PDF. We repeat this procedure 1000 times and compute the mean binned PDF, $\mathbf{d_{\rm G}}$, and its covariance $\Sigma$.

\subsection{PDF of the high-redshift convergence map}
\label{sec:datavecs}

For our actual data vector, we use the convergence map from the Multi-Dark Planck 2 (MDPL2) synthetic sky simulations (Omori in prep.). This simulation suite consisting of realistic maps of various observables such as CMB lensing and galaxy density maps, based on the MDPL2 $N$-body simulation \citep{mdpl}. MDPL2 is a 1 Gpc/h dark matter only simulation with $3840^3$ particles. Unlike the Gaussian simulations described in Section~\ref{sec:gaussian_sim}, these are built on an $N$-body simulation, which contains realistic structure formation and thus non-Gaussian information. 

The CMB lensing map is generated by first projecting all the particles onto HEALPix shells of $N_{\rm side}=8192$ and width 25 Mpc/h, and by running the GRayTrix raytracing code \citep{hamana2015,shirasaki2015} through the stacked density shells. The raytracing is performed up to $z=8.6$, and an additional Gaussian convergence component is added to take into account for the contribution in the redshift range $8.6<z<1100$ to produce the full CMB lensing map.

The galaxy density maps are generated using the same dark matter density shells that were used for raytracing to produce the CMB lensing map, and hence are correlated.
The density shells are weighed according to the $n(z)$ of the LSST-Y1 DESC SRD lens galaxy samples and integrated over the full redshift range to produce mock galaxy density maps of LSST-Y1 galaxies.

One limitation of using this $N$-body simulation is the number of realisations available: since the computational cost of generating multiple $N$-body realisations is prohibitive, only one realisation exists. To circumvent this issue, we generate semi-independent realisations of the observed PDF by adding to the ``true'' data vector contribution from noise and cosmic variance estimated from Gaussian simulations. That is, we have for each realization $i$, the observed PDF data vector
\begin{equation}
{\bf d }_{{\rm NG},i}^{\rm obs} = {\bf d}_{\rm NG}^{\rm true} + \Delta {\bf d}_{i},
\end{equation}
where ${\bf d}_{\rm NG}^{\rm true}$ is the true data vector derived from the mean of the PDFs measured from the full-sky CMB lensing map with 100 independent realisations of CMB lensing noise (which we will refer to as non-Gaussian PDF here after). $\Delta {\bf d}_{i}$ is derived from the Gaussian simulations described in Section~\ref{sec:gaussian_sim}, where   
\begin{equation}
\Delta {\bf d}_{i}\equiv {\bf d}_{{\rm G},i}-\langle {\bf d}_{\rm G} \rangle.
\end{equation}
Here, ${\bf d}_{{\rm G},i}$ is the PDF data vector for one Gaussian simulation and $\langle {\bf d}_{\rm G} \rangle$ is the average over 1000 simulations. We note that here we have assumed that $\Delta {\bf d}_i$ derived from the Gaussian simulations is close to that in the non-Gaussian simulations. This should be a decent approximation given that the differences between Gaussian PDFs and non-Gaussian PDFs are small.

\section{Results}\label{sec:results}

\subsection{Example with ideal case}\label{sec:ideal_case}
Prior to carrying out forecasts of detectability for the individual experiments, we conduct our analysis under an idealized setup 
to gain insights on the hypothetical upper limit of our signal-to-noise ratio.
This is done by assuming a CMB experiment with no noise and full sky coverage. We further construct a noiseless low-redshift lensing template assuming a galaxy experiment with infinite galaxy number density with the same $n(z)$ as LSST-Y1 lens sample, and subtract that off from the CMB lensing map.

\begin{figure}
    \centering
    \includegraphics[width=\linewidth]{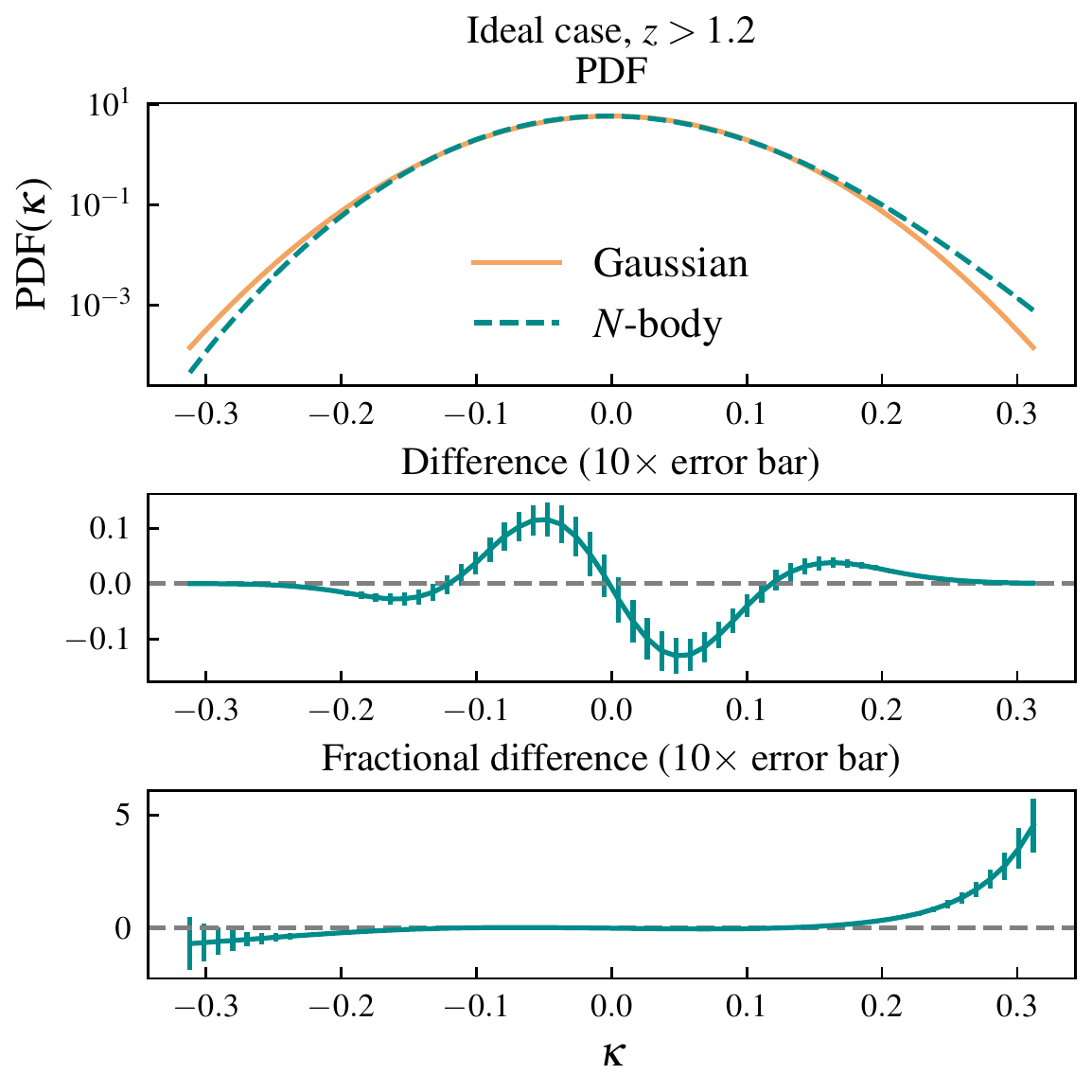}
    \caption{\textit{Top panel:} the comparison between the PDF from non-Gaussian and Gaussian simulations. \textit{Middle and Bottom panels:} the absolute and fractional difference between the non-Gaussian and Gaussian PDFs. The fractional difference is defined as ${\rm PDF}_{\rm non-Gauss}(\kappa)/{\rm PDF}_{\rm Gauss}(\kappa)-1$. The error bars are scaled up by a factor of 10. }
    \label{fig:ideal_hist}
\end{figure}

In the top panel of Figure~\ref{fig:ideal_hist}, the teal line represents the non-Gaussian PDF measured from the $N$-body simulation, with error bars estimated by taking the variance of the PDF over 1000 Gaussian simulations under the same ideal assumptions. The orange line represents the mean PDF of the 1000 Gaussian simulations. The bottom two panels show the absolute and fractional difference between the teal and the orange lines. The comparison of the PDF is consistent with our intuition: the values of $\kappa$ are slightly skewed toward higher values relative to zero, corresponding to the physical picture that overdensities grow faster than underdensities. 

We compute the $\chi^2$ between the non-Gaussian PDFs (${\bf  {d}}_{\rm NG}$) and the mean Gaussian PDF ($\langle {\bf  {d}}_{\rm G}\rangle$), and find 
a mean $\chi^2$ of  $\sim2\times10^4$, which exceeds the number of degrees of freedom  ($\nu=60$). The signal-to-noise ratio in this case is computed to be approximately $146\sigma$. This high $S/N$ value suggests that, in theory, there is significant non-Gaussian information in a CMB lensing maps beyond $z=1.2$.

\subsection{Projected CMB experiments at $z>1.2$}\label{sec:results_main}
\begin{figure*}
    \centering
    \includegraphics[width=\linewidth]{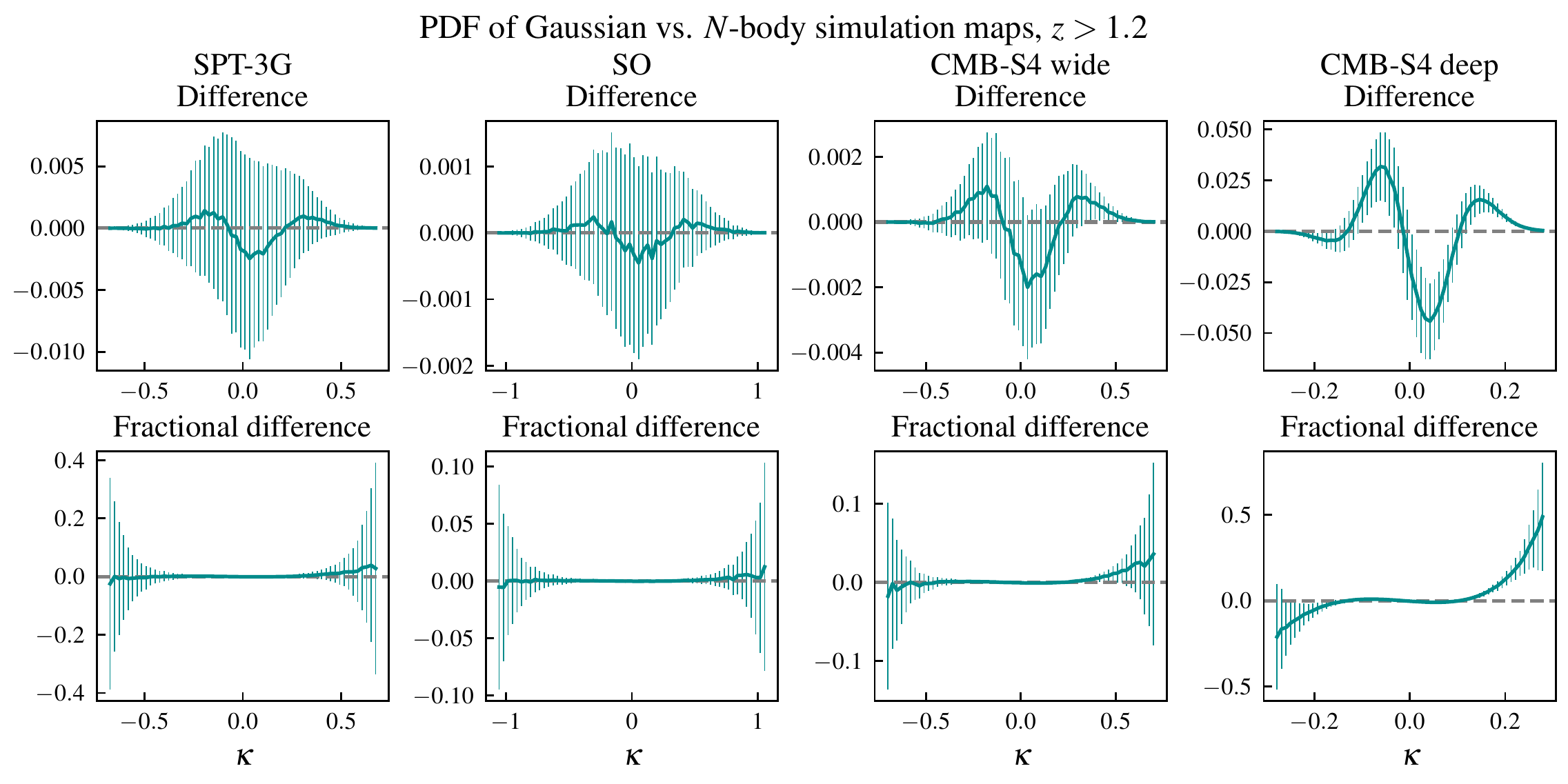}
    \caption{The absolute (upper panels) and fractional (lower panels) difference 
    between the PDF of the non-Gaussian simulations and the
    mean PDF of 1000 Gaussian realizations. Error bars are represented by the vertical lines.}
    \label{fig:histograms}
\end{figure*}

\begin{figure*}
    \centering
    \includegraphics[width=\linewidth]{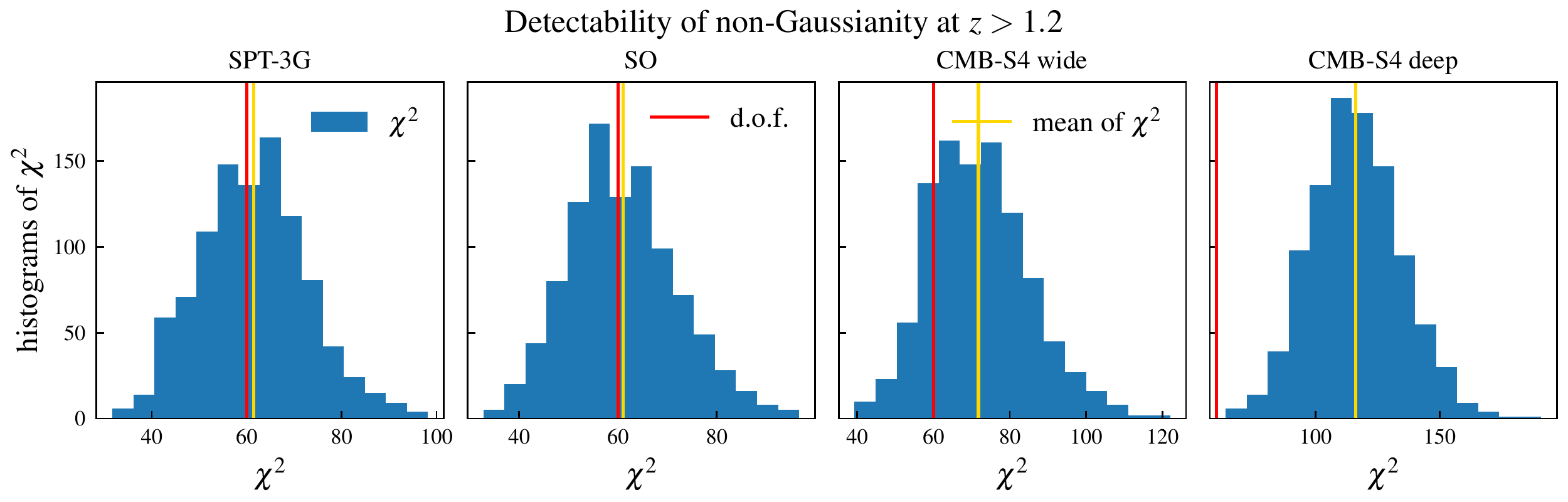}
    \caption{The measured $\chi^2$ between the mean Gaussian PDF and non-Gaussian PDF, for the four CMB experiment configurations.}
    \label{fig:chisquares}
\end{figure*}

In this section, we run our forecasts with more realistic experimental setups as described in Section~\ref{sec:CMB_surveys}. There are two factors that will reduce the signal-to-noise ratio of our detection: 1) CMB noise broadens the PDFs by convolution with a Gaussian kernel and consequently reduces the difference between the Gaussian and non-Gaussian PDF shapes, and 2) since experiments only observe a fraction of the sky, we get extra contributions to the total error budget from the sample variance. For our forecasts, however, we do not include other systematics besides CMB and shot noise.

In  Figures~\ref{fig:histograms} and \ref{fig:chisquares}, we show both the absolute and fractional differences between PDFs measured from Gaussian and non-Gaussian maps, as well as the measured $\chi^2$ distributions measured between Gaussian and non-Gaussian simulations. The results can be summarized as follows:
\begin{itemize}
    \item \textbf{SPT-3G}
    The difference between Gaussian and non-Gaussian PDF is  insignificant when the statistical uncertainties are considered. This is accurately reflected by the signal-to-noise ratio: the mean of the $\chi^2$ is very close to the degree of freedom (61.4 and 60 respectively), yielding a signal-to-noise of 1.2.
    \item \textbf{SO-goal:} 
    From the measured $\chi^2$,  we deduce a smaller signal-to-noise ratio (S/N$\sim1.0$) than the case for SPT-3G.  While we expect that the increased survey area will improve detection, the comparison between SO and SPT-3G indicates that the signal-to-noise ratio scales faster with noise level compared to sky area.
    \item \textbf{CMB-S4 wide}:
    The benefits of a larger area manifest in the signal-to-noise ratio as it achieves S/N$=3.4$ --- although still not a significant detection, it is higher than SPT-3G and SO. 
    \item \textbf{CMB-S4 deep:} Despite the small survey area, the deep field of CMB-S4 yields the most significant detection (S/N$=7.5$) due to the low noise level. The difference in the PDF can also be identified by eye as shown in Figure~\ref{fig:histograms}.
\end{itemize}

In summary, we find that for ongoing or near-term CMB experiments (SPT-3G and SO), detecting deviations from Gaussianity in high-redshift lensing maps would be challenging, and the same conclusion holds for the CMB-S4 wide field survey ($\sim 3.4\sigma$). Only when we switch to the CMB-S4 deep survey, we could expect to find a significant detection ($\sim7.5\sigma$ assuming no systematic uncertainties). We conclude that while the sky area plays an important role in the detection significance, it is more optimal to use deep low-noise experiments for this scientific purpose.

As noted before, these results are obtained without assuming any systematic effects. However, a few systematic effects, including galaxy bias, could limit the accuracy of the removal of the low-redshift contribution to the CMB lensing signal. We discuss in Appendix~\ref{sec:galaxy_bias} the sensitivity of the results in this paper to these potential systematic effects. Another caveat in these results is that the cosmology we use for the Gaussian simulation is exactly the same as the fiducial cosmology for the \nbody{} simulation. In reality, however, the true cosmology is not known precisely; any offset in the input cosmology for the Gaussian simulations will also result in a discrepancy of the level of non-Gaussianity.

\subsection{Redshift dependence}
\begin{figure}
    \centering
    \includegraphics[width=\linewidth]{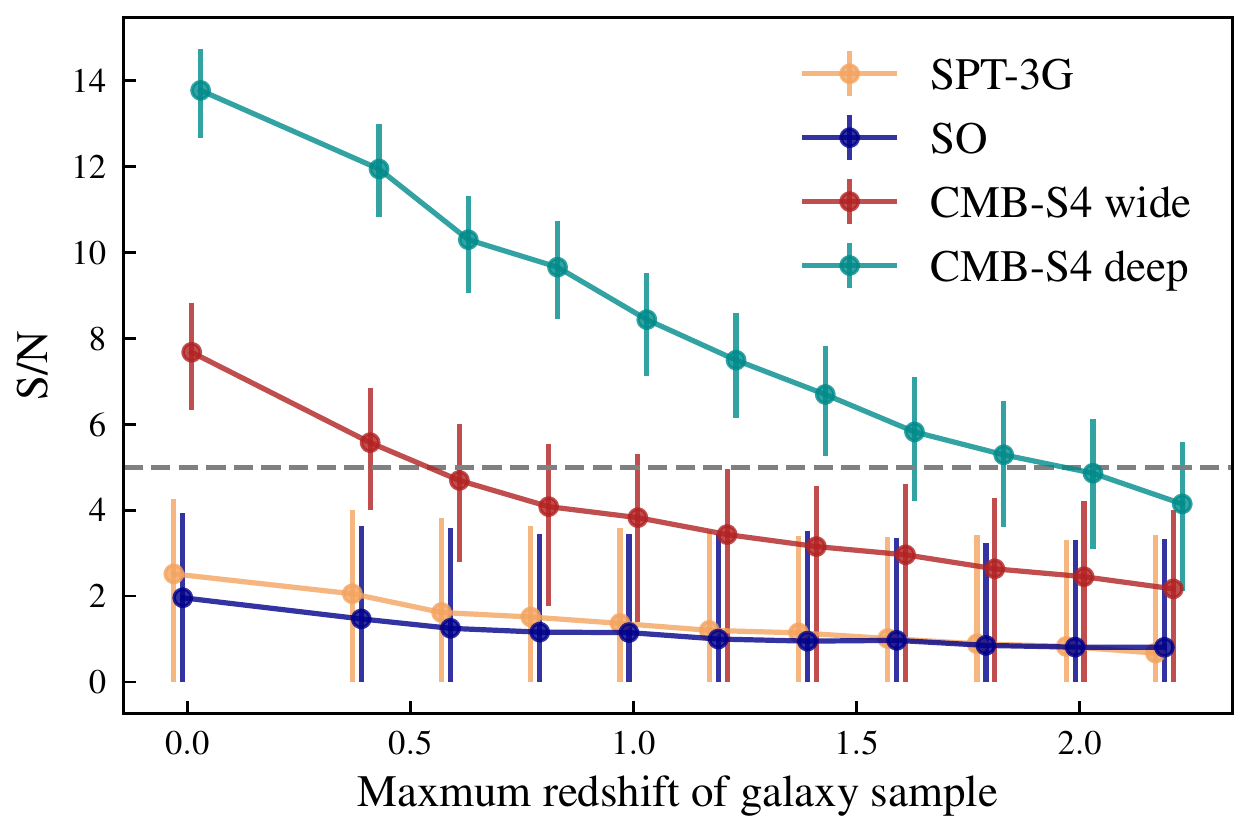}
    \caption{The mean signal-to-noise vs. the maximum redshift of the galaxy sample for the four CMB experiments. The error bars indicates the range of the central 66\% of signal-to-noise ratios. The dashed grey line corresponds to $S/N=5$, which we consider to be the threshold of significant detection. }
    \label{fig:redshift}
\end{figure}

\begin{figure}
    \centering
    \includegraphics[width=\linewidth]{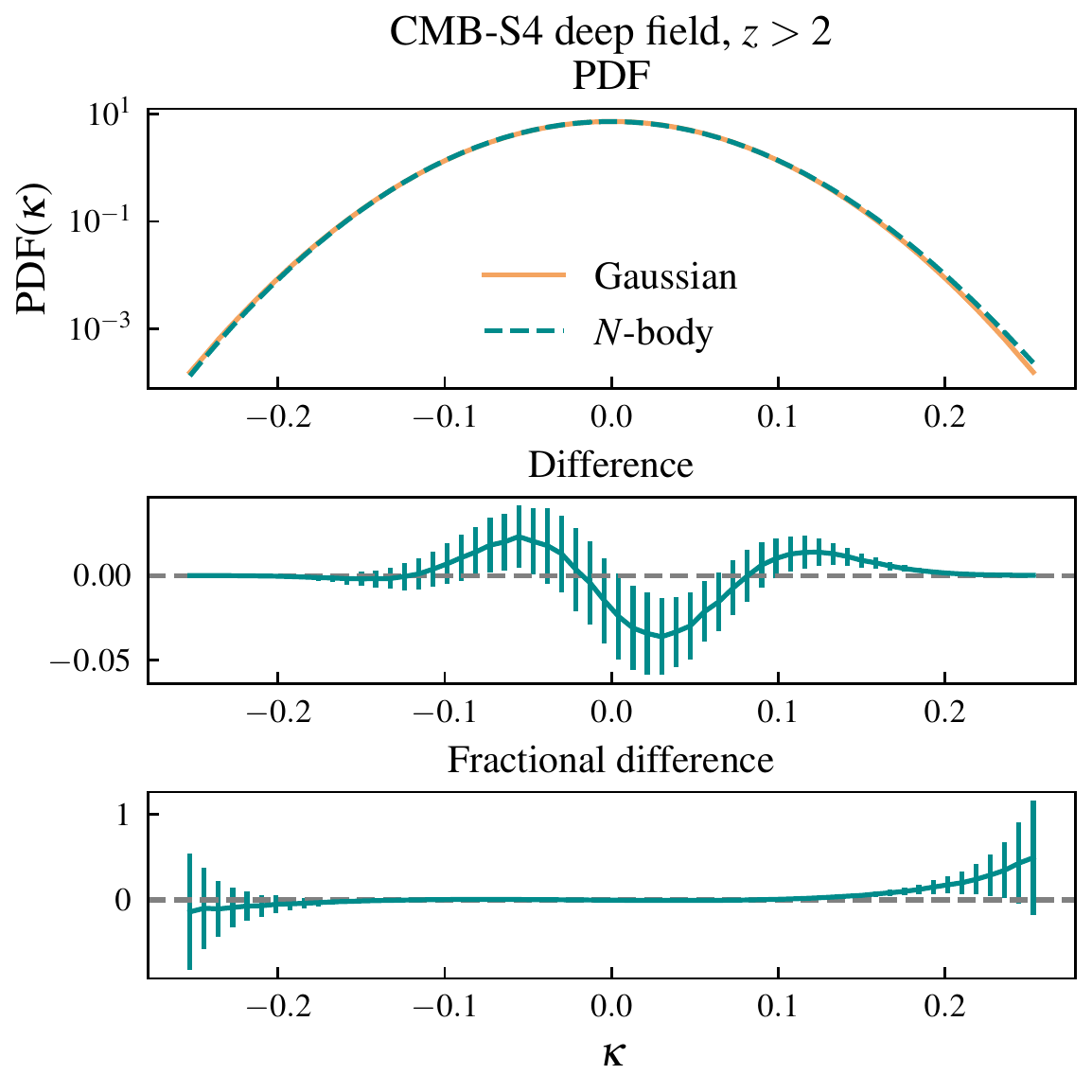}
    \caption{The detected deviation of PDF from a Gaussian distribution for the CMB-S4 deep field at $z>2$. This is the highest redshift above which we can obtain a significant detection ($5\sigma$) with CMB-S4 deep field.}
    \label{fig:maxdetection}
\end{figure}

In this section, we investigate whether the maximum redshift of the lens galaxy sample we use to create the low-redshift template affects the detectability of non-Gaussianity.

We repeat the computation in Section \ref{sec:results_main}, but using different numbers of redshift bins to remove the low-redshift contribution. Starting from the original lensing map (i.e. no removal), we progressively remove the low-redshift contribution up to $z=2.2$. We plot the signal-to-noise ratio of the resulting non-Gaussian signal in Figure~\ref{fig:redshift}. 

Several conclusions can be drawn from our results. First, the signal-to-noise ratio of detecting non-Gaussian information is the highest for the original CMB lensing maps, and decreases as we remove more contributions from low redshifts. This is consistent with our intuition that non-Gaussianity is most prominent at the lowest redshifts. Second, as in the fiducial cases discussed in Section \ref{sec:results_main}, both SO and SPT-3G yield insignificant detections. This is true even in the case of using the full CMB lensing map without any low-redshift removal. We find a higher signal-to-noise for the CMB-S4 wide survey but only up to $z \sim 0.4$. The CMB-S4 deep survey provides the most significant detection of non-Gaussianity, and the signal-to-noise ratio remains above 5 up to $z \sim 2.0$. The measured PDFs for the case of removing up to $z=2$ is shown in Figure~\ref{fig:maxdetection}). This suggests that applying our methodology to the CMB-S4 deep survey data and LSST-Y1 galaxies, we will be able to reconstruct the history of the nonlinear structure growth from $z\sim2$ to $z=0$. 

Figure~\ref{fig:redshift} is also interesting for two other reasons. First, it provides a ``lookup'' table for the expected signal-to-noise of the non-Gaussian signal from a $\Lambda$CDM Universe as a function of redshift. Given the science interest, one can then choose different configurations of the analysis to optimize the outcome. For example, from the plot we know that if we want the highest signal-to-noise cosmological constraint from all the non-Gaussian signal from CMB-S4 deep we would not remove any low-redshift contribution, but if we want that constraint from the $z>0.5$ Universe we will still have a considerable level of signal-to-noise. Second, we can also potentially use this plot as a data vector to study the redshift-evolution of non-Gaussianity. That is, we can make a measurement such as Figure~\ref{fig:redshift} on future observational data and compare with simulations; any significant deviations may hint towards something interesting -- either our incomplete understanding of how structures form over the cosmic history or the existence of non-Gaussianity in matter distribution in the early Universe.

\section{Conclusion} \label{sec:conclusion}

In this work, we explore the potential of detecting deviations from Gaussianity in the $z>1-2$ Universe by looking at high-redshift mass maps. To do this, we utilize CMB lensing maps, which probe all the matter between us and the last scatter surface, and apply a procedure to remove the $z<1-2$ lensing contribution using data from galaxy surveys. We perform a simulated analysis using mock CMB lensing maps and galaxy samples from ongoing/future CMB (SPT-3G, SO, CMB-S4) and galaxy (LSST) surveys. Being able to detect non-Gaussianity at these intermediate to high redshifts allows us to study the early stages and the history of structure formation.

We find that, in theory, there is significant non-Gaussian information above $z=1.2$ (roughly the redshift range probed by LSST-Y1 lens galaxies) in the CMB lensing maps due to structure growth. However, with most CMB experiments we have considered (SPT-3G, SO, and CMB-S4 wide), this signal is not detectable due to the noise levels in the maps. We do find, however, that the CMB-S4 deep survey is likely to detect deviations from Gaussianity at $\sim7.5\sigma$ at $z>1.2$  in the absence of systematics. We also explore the possibility of using a galaxy sample out to $z=2$, and find that we sould expect to see signatures of deviation from Gaussianity at $\sim5\sigma$.

There are a number of simplifications we have made in this analysis. Firstly, we have used Gaussian simulations to approximate the non-Gaussian covariance. At the level of noise adopted in this paper, this is a good approximation and should not qualitatively change our conclusions. Secondly, to obtain the ``true'' PDF data vector, we have ignored the effect of cosmic variance. Thirdly, we have not removed the lensing contribution from $0<z<0.2$ and therefore it is included in the non-Gaussian signals. However, we expect this contribution is very small (since the kernel of CMB lensing falls steeply below $z=0.5$ and therefore does not contribute to the overall CMB lening signal) and can be ignored. Finally, we note that shown in Appendix~\ref{sec:galaxy_bias}, the non-Gaussianity measurement in high-redshift mass maps may require a precise understanding of some nuisance parameters such as galaxy bias. In this work, we have not assumed any observational systematic effects in either the galaxy sample (photometric redshift, masks, survey inhomogeneity) or the CMB lensing map (foreground, filtering of a given survey), which would could in principle further degrade the signal, but the inclusion of these effects is outside the scope of this paper.

There are several directions one can explore in more detail following this work. First, one can study these non-Gaussianities using alternative summary statistics such as three point correlation functions or moments of the PDF, following the framework of this paper. These alternative statistics may have practical advantages over PDFs (e.g. less sensitivity to noise and systematic effects). Second, it will be crucial to incorporate effects such as photometric redshift uncertainty, foreground contamination, and baryonic feedback to come up with a more realistic forecast of the detectability of non-Gaussianity signals at high-redshifts.

\section*{Acknowledgement}

The authors would like thank Eric Baxter for reviewing and commenting on the draft. 

ZZ, YO and CC are supported by DOE grant DE-SC0021949.

We gratefully acknowledge the computing resources provided on Crossover, a high-performance computing cluster operated by the Laboratory Computing Resource Center at Argonne National Laboratory.

This research used resources of the National Energy Research Scientific Computing Center, which is supported by the Office of Science of the U.S. Department of Energy under Contract No. DE-AC02-05CH11231. 

\section*{Data Availability}
No new data were generated or analysed in support of this research.

\bibliography{ref}

\appendix

\section{Validation of the recovered high-redshift CMB lensing map}
\label{sec:validation}
\begin{figure}
    \centering
    \includegraphics[width=0.45\textwidth]{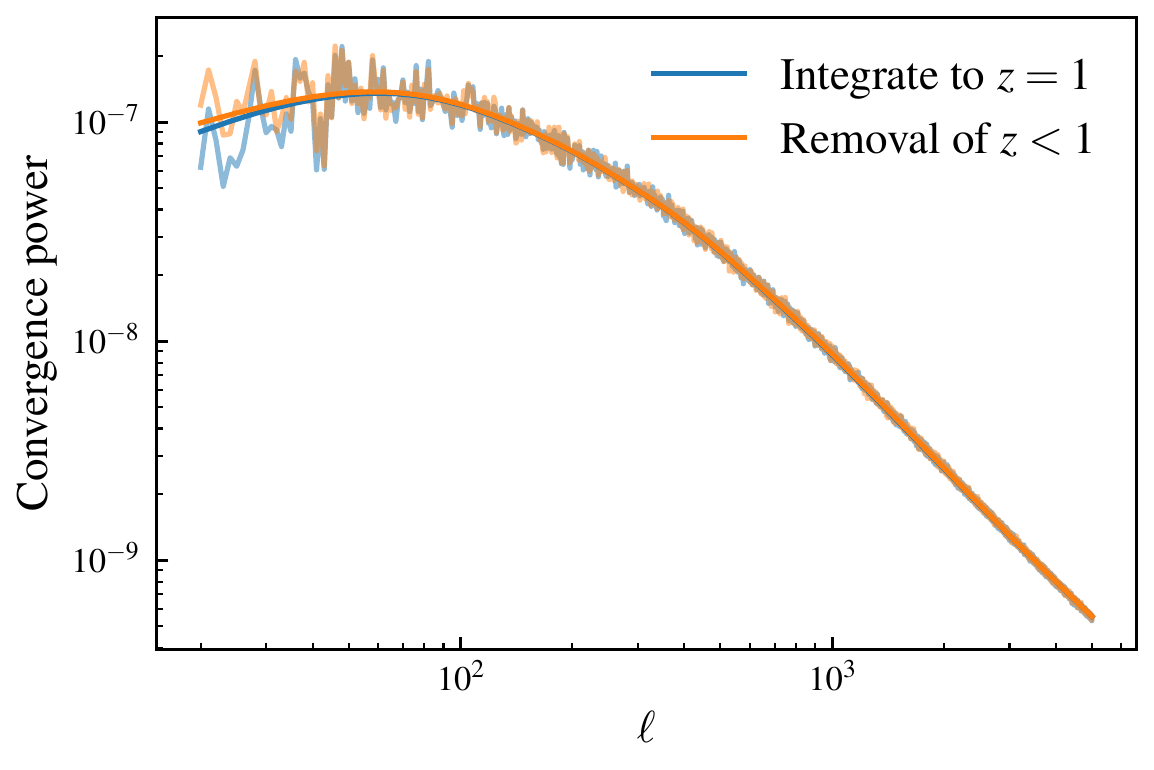}
    \caption{Comparison between the power spectra of the $N$-body simulation integrated down to $z=1$ (blue) and a recovered high-redshift map above $z=1$ using the method described in Section~\ref{sec:delens} (orange). The solid lines are predicted power spectra from theory and transparent lines are measured power spectra from a map realization.}
    \label{fig:verify_delens}
\end{figure}

One way to verify the recovered high-redshift CMB lensing map using the method described in Section~\ref{sec:delens} is to compare it with the $N$-body simulation that only integrates the CMB lensing convergence down to a certain redshift (e.g. $z=1$). The power spectra measured on the recovered map and the $N$-body simulation map are expected to match. Figure~\ref{fig:verify_delens} shows a comparison between the power spectra of the recovered and $N$-body simulation maps, where we use five top-hat $n(z)$s with equal width of 0.2 between $z=0$ and $z=1$. We restrict $\ell$ from 30 to 5000, where the filter described in Section~\ref{sec:filter} is typically non-zero. Within this range, the two power spectra agree well with each other.

\begin{figure*}
    \centering
    \includegraphics[width=\textwidth]{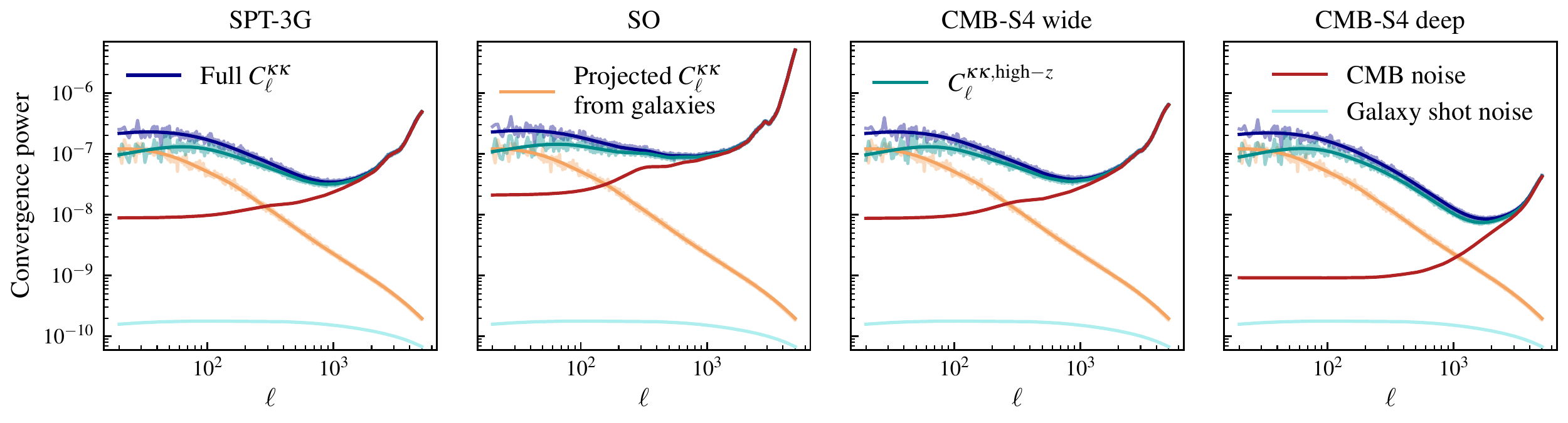}
    \caption{The same as Figure~\ref{fig:delens_ideal}, but the full and high-redshift CMB lensing power spectra now include the projected CMB noise, which are also shown (red).}
    \label{fig:cls}
\end{figure*}

\section{High-redshift CMB lensing power spectrum}\label{sec:cls}
Similar to Figure~\ref{fig:delens_ideal}, Figure~\ref{fig:cls} shows the power spectra of the full CMB lensing map, projected CMB lensing map from low-redshift galaxies, the high-redshift CMB lensing map, and effective galaxy shot noise for the CMB lensing noise levels of the experiments we consider in this study. The signal-to-noise ratio is represented by the difference between the amplitudes of $C_\ell^{\kappa\kappa,\text{high-}z}$ (teal) and CMB noise (red). One can visually compare the signal and noise levels across the CMB experiments. Galaxy shot noise is small  throughout all four cases, and therefore has minimal impact on the level of detection.

\section{Sensitivity to systematic effects}\label{sec:galaxy_bias}
\begin{figure}
    \centering
    \includegraphics[width=0.48\textwidth]{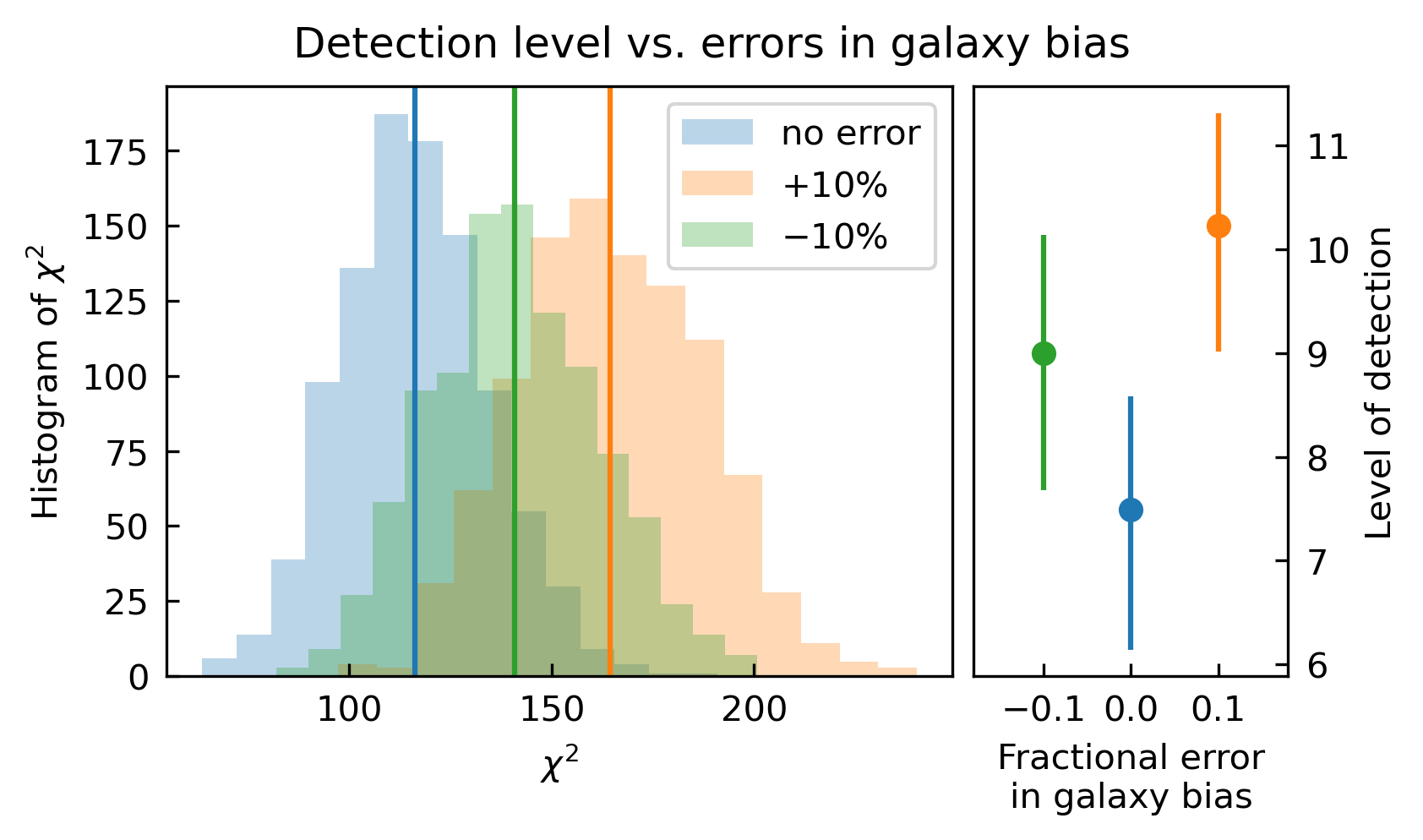}
    \caption{This plot shows how errors in galaxy bias impacts the detectability of non-Gaussianity. \textit{Left panel}: the blue histogram shows the \(\chi^2\) statistics if the galaxy bias is perfectly known. The green and orange histograms show the \(\chi^2\) statistics if the galaxy bias is under/over-estimated by \(10\%\). The vertical lines represent the mean of \(\chi^2\) for each case. \textit{Right panel}: the signal-to-noise ratio of non-Gaussian signatures in PDF($\kappa$) for each scenario. }
    \label{fig:galaxy_bias}
\end{figure}

The removal of low-redshift signals from CMB lensing maps may be subject to a number of systematics such as photometric redshift uncertainty and galaxy bias. In this Appendix, we assess their impact on our main analysis.

As an example, we investigate how our main result will change if we have mis-estimated the galaxy bias of our sample. We look at the CMB-S4 deep field case at redshifts $z>1.2$ as in the right panel of Figure~\ref{fig:chisquares}. Specifically, we reproduce our analysis with a $\pm 10\%$ offset in linear galaxy bias and study the effects on the $\chi^{2}$ distribution as shown in Figure~\ref{fig:galaxy_bias}. Interestingly, both under- and overestimation of the galaxy bias increase the detection significance. As shown in the right panel of Figure~\ref{fig:galaxy_bias}, a 10\% error on galaxy bias can cause an overestimation of the signal-to-noise ratio by roughly one to two standard deviations. This result highlights the importance of having an accurate estimate of galaxy bias. Fortunately, we expect this would be possible in future galaxy surveys by combining galaxy clustering and weak lensing \cite{prat2023}. 

One should also note that, according to Eq.~\ref{eq:delens_alm}, the removal of low-redshift lensing signal only involves measuring cross power spectra from the observed galaxy density fields and the CMB lensing field. Therefore, the obtaining of high-redshift mass map itself is independent of galaxy bias. In our analyses, galaxy bias affects the level of non-Gaussianity only through the Gaussian simulations against which we compare our measurement. Our simulations for PDF($\kappa$) is particularly sensitive to galaxy bias since the amplitude of the high-redshift lensing convergence map is directly proportional to the galaxy bias. Other summary statistics could potentially be less affected by this error.

\end{document}